\documentclass[aps,showpacs,a4paper]{revtex4}
\usepackage{graphicx}

\def\ket{\rangle}
\def\<{\langle}
\def\>{\rangle}

\begin{document}

\title{No-relationship between impossibility of faster-than-light quantum
communication and distinction of ensembles with the same density
matrix\footnote{Accepted for publication in Communications in
Theoretical Physics}}
\author{Chuan Wang$^{1}$, Gui Lu Long$^{1,2}$ and Yang Sun$^{3,1}$ }
\address{$^1$ Department of Physics and Key Laboratory For Quantum
Information and Measurements\\ Tsinghua University, Beijing
100084, People's Republic of China\\
$^2$ Key Laboratory for Atomic and Molecular NanoSciences\\
Tsinghua University, Beijing 100084, People's Republic of China\\
$^3$ Department of Physics, University of Notre Dame, Notre Dame,
Indiana 46556, USA }
\date{\today }

\begin{abstract}
It has been claimed in the literature that impossibility of
faster-than-light quantum communication has an origin of
indistinguishability of ensembles with the same density matrix. We
show that the two concepts are not related. We argue that: 1) even
with an ideal single-atom-precision measurement, it is generally
impossible to produce two ensembles with exactly the same density
matrix; or 2) to produce ensembles with the same density matrix,
classical communication is necessary. Hence the impossibility of
faster-than-light communication does not imply the
indistinguishability of ensembles with the same density matrix.
\end{abstract}

\maketitle

The study on quantum information and quantum computation has
become one of the research focuses worldwide. The combination of
quantum mechanics with information and computer sciences has
produced many fruitful results, which may become an advanced
technology in the future. A quantum computer can offer additional
computing power which can greatly speed up the solution for a
prime factorization of a large number \cite{r1}, and for an
unsorted database search \cite{r2}. The quantum key distribution
offers unconditional security in secret communication \cite{r3}.
With quantum entanglement, users at distant sites may share
particles that are part of an entangled system to fulfill certain
communication task, as for instance in sharing a secret \cite{r4}.
Thus it is tempting to look for more applications of quantum
mechanics.

One such search is the faster-than-light communication using
shared entangled particles. However, any faster-than-light motion
is in an obvious violation of special relativity. By using
ensembles of qubits that are parts of entangled pairs, a scheme
was proposed in Ref. \cite{r5} to show that a faster-than-light
communication is impossible. The impossibility is claimed to
relate to the indistinguishability of ensembles with the same
density matrix \cite{r5,r6}. On the other hand, however, it has
recently been pointed out that ensembles with the same density
matrix can be distinguished physically \cite{r8,r7}, and this
conclusion is closely linked to the quantum nature of NMR quantum
computing \cite{r9}. There is apparently an contradiction between
these two results.

One uses Einstein-Podolsky-Rosen (EPR) pairs separated in space as
means of communication. Then a measurement of two different
observables collapses the EPR pairs and produces two ensembles
having the same density matrix. The two different measurements
transmit one bit of information if one can distinguish these two
ensembles having the same density matrix. In this paper, we will
show that the conclusion of distinction of ensembles with the same
density matrix bears no relationship with the impossibility of
faster-than-light communication. We show that there is a flaw in
Preskill's argument in Ref.\cite{r5}: the two ensembles produced
by measuring EPR pairs using either $\sigma_x$ or $\sigma_z$ bases
will generally NOT produce ensembles with identical density
matrix. The conclusion drawn in Ref.\cite{r5} should be rephrased
into: it is impossible to produce two ensembles with the same
density matrix by measuring particles from $N$ EPR pairs without
classical communication.

\begin{figure}[tbp]
\begin{center}
 \includegraphics[width=8cm,angle=0]{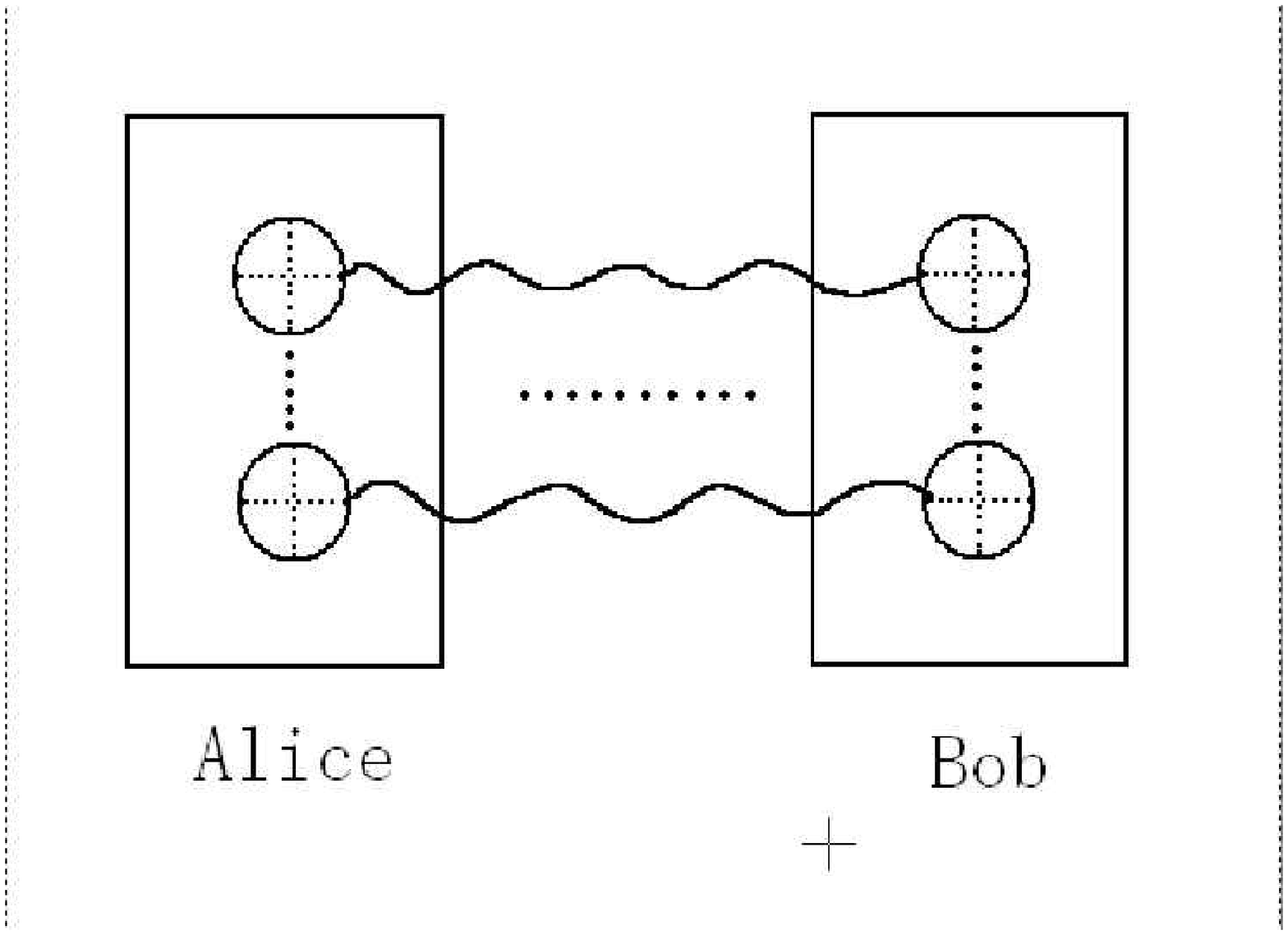}
\end{center}
\caption{ Each pair of qubits connected by a wave line represents
an EPR pair. } \label{f1}
\end{figure}

\begin{figure}[tbp]
\begin{center}
 \includegraphics[width=8cm,angle=0]{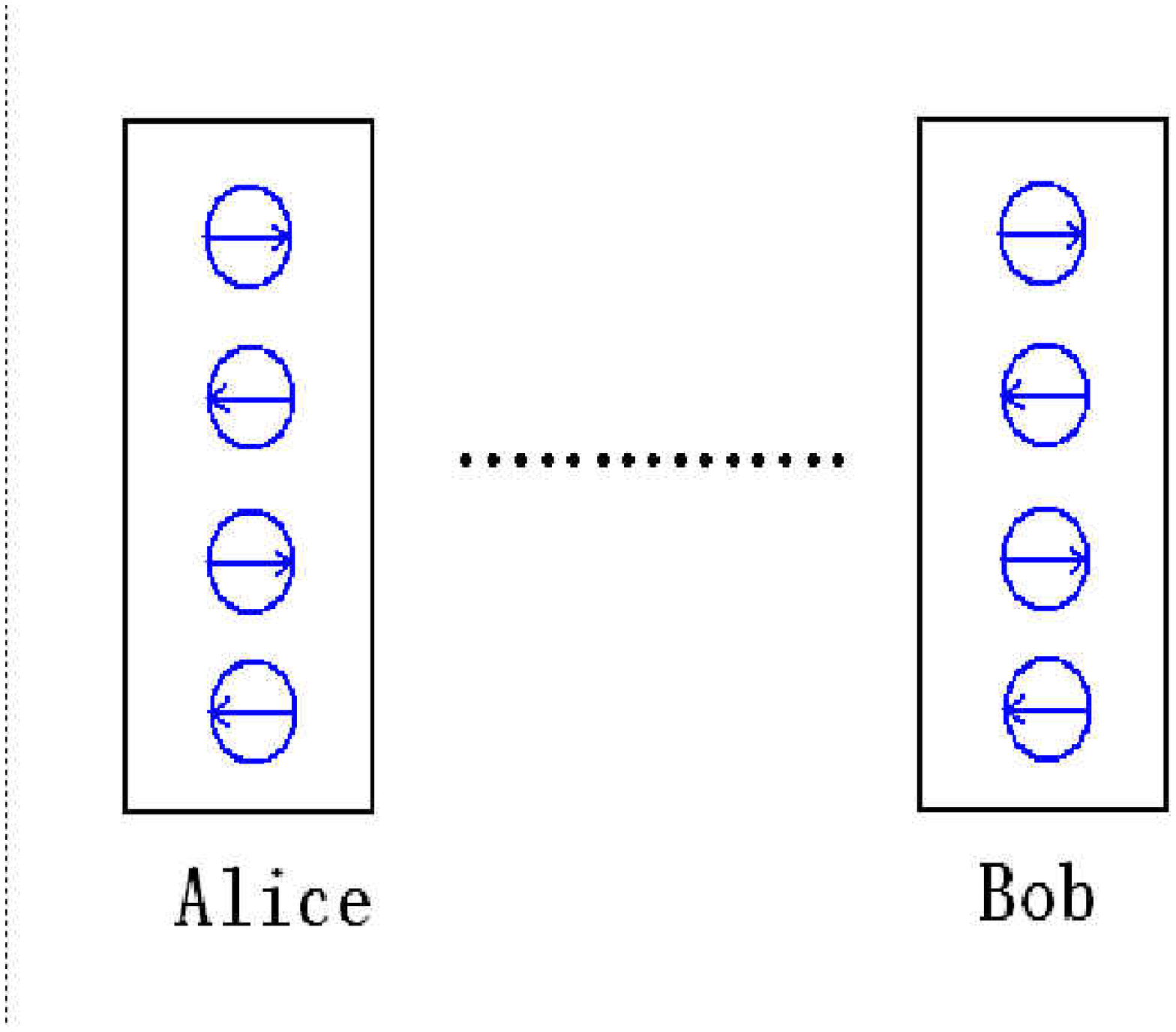}
\end{center}
\caption{ By measuring $\sigma_{x}$ on each qubit, each pair
collapses into $|\uparrow_x\ket$ or $|\downarrow_x\ket$. This
action  represents the value 0.} \label{f2}
\end{figure}

\begin{figure}[tbp]
\begin{center}
 \includegraphics[width=8cm,angle=0]{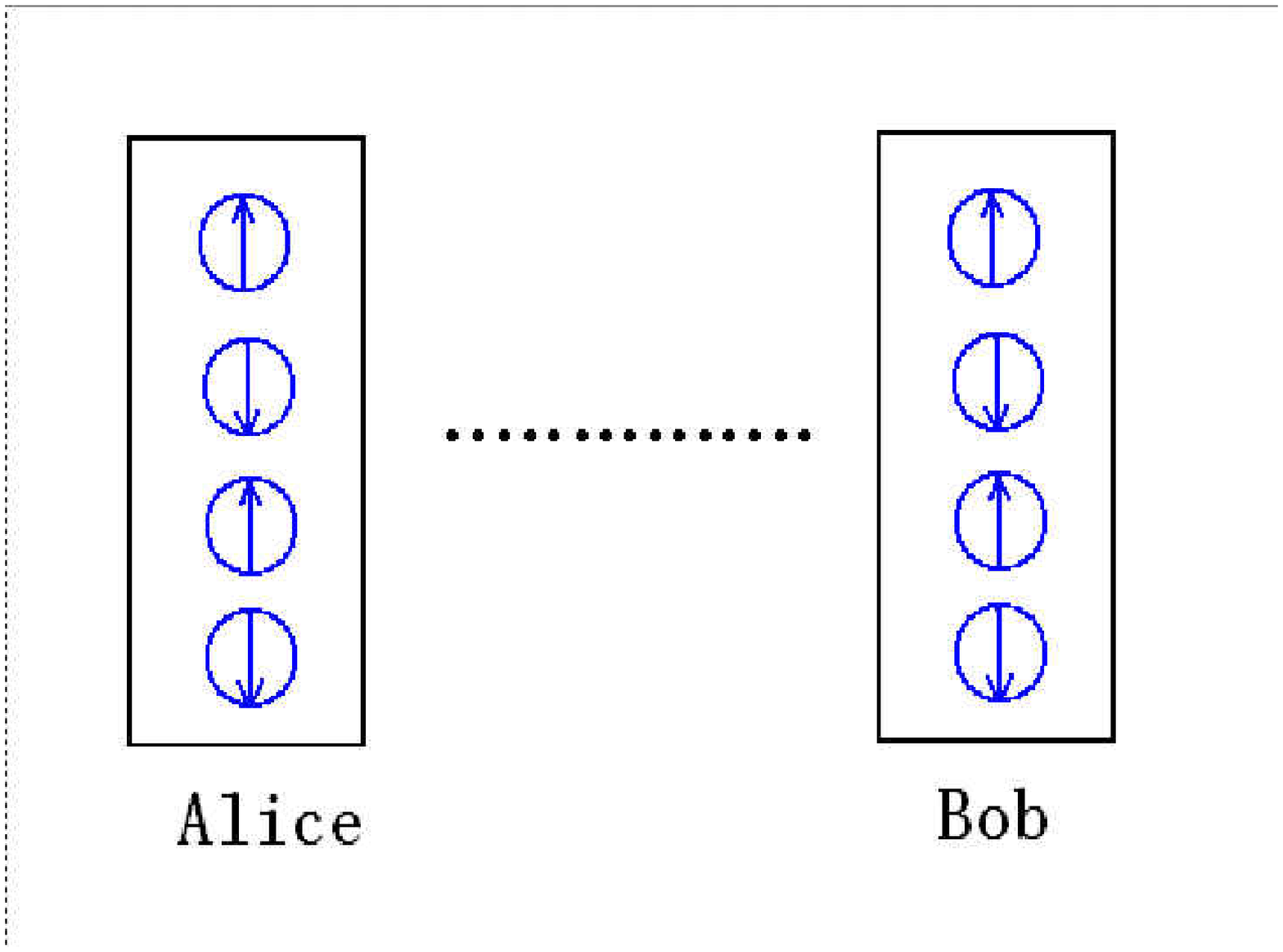}
\end{center}
\caption{  By measuring $\sigma_{z}$ on each qubit, each pair
collapses into $|\uparrow_z\ket$ or $|\downarrow_z\ket$. This
action, the $\sigma_z$ measurement, represents the value 1.}
\label{f3}
\end{figure}

{\it\bf First we briefly review the Preskill scheme}. As shown in
Fig. \ref{f1}, Alice and Bob share $N$ pairs of qubits in state
\begin{equation}
|\Phi_{AB}\rangle=\frac{1}{\sqrt{2}}(|\uparrow_{Z}\rangle_{A}|\uparrow_{Z}
\rangle_{B}+|\downarrow_{Z}\rangle_{A}|\downarrow_{Z}\rangle_{B}).
\label{e1}
\end{equation}
Alice and Bob are separated by a large distance. Bob could send
Alice a one-bit message by measuring his particles with either
$\sigma_{x}$ as shown in Fig.\ref{f2} or $\sigma_{z}$ as shown in
Fig.\ref{f3}, thus preparing Alice's spins in either
($|\uparrow_{Z}\rangle_{A},|\downarrow_{Z}\rangle_{B}$) or
($|\uparrow_{X}\rangle_{A},|\downarrow_{X}\rangle_{B}$). The two
ensembles produced by these $\sigma_x$ or $\sigma_z$ measurement
have the same density matrix, \begin{eqnarray} \rho_A=\left(
\begin{array}{cc} {1\over 2} & 0\\
                   0      & {1\over 2}\end{array}\right).
                   \end{eqnarray}
If Alice could tell the difference between these two ensemble
preparations, then she would be able to read Bob's information
immediately, accomplishing a faster-than-light communication.


However, this is impossible. Preskill stressed in Ref. \cite{r5}
that ``though the two preparation methods are surely different,
both ensembles are described by precisely the same density matrix
$\rho_{A}$. Thus there is no conceivable measurement Alice can
make that will distinguish the two ensembles, and no way for Alice
to tell what action Bob performed. The `message' is unreadable".

Preskill then provided a method for Alice and Bob to distinguish
the two ensemble preparations with some additional help. Alice can
choose a measuring device, say $\sigma_{x}$, to measure each of
her particles and compare with the result of Bob which was
transmitted to him through a telephone line. If her results have
perfect agreement with Bob, then she knows that Bob's action is
$\sigma_{x}$, otherwise Bob's action is $\sigma_{z}$. Here ideal
conditions are assumed for simplicity. Apparently, the two
ensemble preparations can be distinguished. However, Preskill
stressed that faster-than-light communication is not possible
because telephone calls are needed for the distinction, and signal
in a telephone line travels at the speed of light.

Preskill has related the impossibility of faster-than-light
communication to the indistinguishability of ensembles having the
same density matrix.

{\it\bf There is a flaw in the Preskill analysis.} Apparently
Preskill's analysis ignored the fluctuation in the measured
result. It is true that in both measurements, each particle has
$1/2$ probability to collapse into $|\uparrow_{Z}\rangle$
($|\uparrow_{X}\rangle$)  or
$|\downarrow_{Z}\rangle$($|\downarrow_{X}\rangle$), but the number
of particles in the $|\uparrow_{Z}\rangle$
($|\uparrow_{X}\rangle$) direction is not exactly the same as that
in the $|\downarrow_{Z}\rangle$($|\downarrow_{X}\rangle$)
direction. Hence the density matrix of one ensemble is
\begin{eqnarray}
\rho_A=\left(\begin{array}{cc} {1\over 2}-{N_\delta \over N} & 0\\
                                 0                 & {1\over 2}+{N_\delta \over
                                 N}\end{array}\right),
                                 \end{eqnarray}
where $N_\delta$ is a random number that could be positive or
negative and is proportional to $\sqrt{N}$. $N_\delta$ indicates
the difference between the number of particles in the
$|\uparrow_{Z}\rangle$ ($|\uparrow_{X}\rangle$) state and the
$N/2$.  We stress that the two different ensemble preparations in
general leads to different density matrices if the number of
particles is finite, which is usually true in real physical
circumstances. Because one can measure the particles in the
ensembles one by one, the absolute value of $N_\delta$ increases
with $N$. Thus as $N$ goes large, the fluctuation becomes large
too. This makes the distinction of the ensembles more easily. Even
if Bob repeats the same kind of measurement, say $\sigma_x$, he
would not be able to produce exactly two identical ensembles if
the particle number is finite. Thus strictly speaking, the density
matrices of the two ensembles produced by $\sigma_x$ and
$\sigma_z$ measurement are not identical.  Though the two
ensembles have different density matrices, one could not use them
for communication as the fluctuation is uncontrollable. Hence
there is no question of distinction of ensembles having the same
density matrix at all in this problem.

{\bf Now we show that even one can distinguish ensembles with the
same density matrix, it is still impossible to perform
faster-than-light communication.} We make the following
modifications to Preskill's scheme. While preparing the ensemble
by measuring each qubit with $\sigma_{z}(\sigma_{x})$, Bob can
make the number of particles in $|\uparrow_{z}\rangle$
($|\uparrow_{x}\rangle$) and
$|\downarrow_{z}\rangle$($|\downarrow_{x}\rangle$) exactly the
same by dropping some qubits. He then tells Alice which qubits
should be excluded from her ensemble, so that Alice's ensemble is
prepared with exactly equal numbers of qubits in opposite
polarization. Of course, Bob and Alice can produce several such
copies with equal or near equal total number of particles. But
they all have equal number of particles in opposite directions. Of
course, the communication in the preparation  is classical, hence
it is not faster-than-light communication.  Now Alice's ensemble
has exactly the density matrix $\rho_A={1\over 2}{\rm I_2}$ and
with two possible form of constituents: either polarized or
anti-polarized along $z$-direction, or polarized or anti-polarized
along $x$-direction. The question is that can Alice distinguish
the two cases with whatever methods that is available. Alice can
determine which one of the two constructions the ensemble is by
making a $\sigma_z$ measurement on each of the particles in the
ensemble. If the ensemble was prepared by the $\sigma_z$
measurement with equal numbers of particles in opposite directions
with the help of classical communication, then the sum of the all
the measured results is zero, namely \begin{eqnarray}
\Sigma_z=\sum_{i=1}^N\sigma_z(i)=0. \end{eqnarray} However if the
ensemble was prepared by the $\sigma_x$ measurement instead, then
the sum of all the measurement will be, \begin{eqnarray}
\Sigma_z'=\sum_{i=1}^N\sigma_z(i)\approx \pm\sqrt{N}.
\end{eqnarray} Because in the first ensemble, the state of the
individual particle  is the eigenstate of $\sigma_z$, and it will
give a definite result when $\sigma_z$ is measured. While in the
second case, the state of an individual particle is in the
eigenstate of $\sigma_x$, there is fluctuations in the measured
results though the average total sum is zero. Because Alice and
Bob have several copies, these fluctuations can be easily found
when Alice repeats the measurement on several such copies. By
observing the fluctuation, Alice can easily determine what
measurement Bob has performed. This has been suggested by
d'Espagnat \cite{r8} and has been generalized into general
ensembles  in Ref. \cite{r7} .

To summarize, it has been shown that ensembles with the same
density matrix can be distinguished physically does not contradict
the claim of impossibility of faster-than-light communication for
two reasons. First, it is impossible to produce ensembles having
the same density by measurements even if a single atom precision
is available. This is because the collapse of state under
measurement is random and the measured results have fluctuations.
This makes the precise density matrix to fluctuate around the
average value. Second, even if two ensembles are produced with
exactly the same density matrix and they are distinguishable by
observing the fluctuations of observables related to the whole
ensemble, this could not be used for information transmission
because the classical communication is necessary to prepare the
ensembles with the same density matrix.

This work is supported by the National Fundamental Research
Program Grant No. 001CB309308, China National Natural Science
Foundation Grants No. 60433050, 10325521, the Hang-Tian Science
Fund, the SRFDP program of Education Ministry of China.

\end{document}